# Zero-Collateral Lotteries in Bitcoin and Ethereum

Andrew Miller  Iddo Bentov
UIUC  Cornell University

## Abstract

We present cryptocurrency-based lottery protocols that do not require any collateral from the players. Previous protocols for this task required a security deposit that is $O(N^2)$ times larger than the bet amount, where $N$ is the number of players.

Our protocols are based on a tournament bracket construction, and require only $O(\log N)$ rounds. Our lottery protocols thus represent a significant improvement, both because they allow players with little money to participate, and because of the time value of money.

The Ethereum-based implementation of our lottery is highly efficient. The Bitcoin implementation requires an $O(2^N)$ off-chain setup phase, which demonstrates that the expressive power of the scripting language can have important implications. We also describe a minimal modification to the Bitcoin protocol that would eliminate the exponential blowup.

## 1 Introduction

The "$N$-player commitment lottery" provides a useful test-case for the expressive power of cryptocurrencies. The rules are simple: each of the $N$ players contributes a quantity of money (say ฿1) into a pot, and one of the players is selected at random to receive the entire reward of ฿$N$.

This lottery game can be desirable as application on its own, but is also studied independently as a useful "leader election" primitive for building randomized distributed systems [1]. Cryptographic protocols for this task based on "commitments" are well-known in the research literature. Though there are many ways of implementing a lottery (such as by using a randomness beacon to draw a random number), a secure implementation of the commitment-based approach is the one with the least reliance on trust assumptions. This is because it can guarantee that an honest player will win the pot with essentially $1/N$ probability, regardless of external factors and the behavior of the other players.

The commitment-based approach can be thought of as relying on a hash function and a public bulletin board. A simple protocol can specify that each party $P_i$ commits to a random value $X_i$ by hashing it (i.e., $h_i = \mathsf{hash}(X_i)$) and posting this hash $h_i$ on the bulletin board. After all the commitments are collected, the parties reveal their $X_i$ values and the winner is determined by combining them, e.g., as $(\sum_i X_i) \bmod N$. For such a protocol to be secure against malicious or rational coalitions, it should handle aborts in a way that ensures that the expected reward of an honest player is never negative.

Cryptocurrencies like Bitcoin and Ethereum provide a natural platform for implementing such lottery protocols, since they can serve both as the public bulletin board and as a way of collecting the initial money deposits and executing the payout procedure. Several works have shown constructions of cryptocurrency-based lotteries with varying performance tradeoffs (e.g., number of rounds versus total communication complexity).

In this work we focus primarily on minimizing the "collateral cost" of commitment-based lottery protocols. In order to guarantee fairness, prior known protocols required each player to deposit an extra amount of money up front (in addition to the ฿1 bet itself), which may be forfeited in the case that that player deviates from the protocol. Previous solutions for Bitcoin-based lotteries (see [4, 9]) had a fairly steep collateral deposit: one that is $O(N^2)$ times larger than the bet amount, *per player*. Even though honest players are guaranteed to get their collateral back at the end, this requirement imposes an "opportunity cost" since the collateral is encumbered for the duration of the protocol; put another way, parties may not be able to afford to borrow the needed collateral.

In this paper we show novel protocols for reducing the collateral costs of $N$-player lotteries, in both Bitcoin (post-SegWit) and in Ethereum, that are the first not to require any collateral deposits whatsoever. Our protocol is based on two components: a) a collateral-free implementation of the simple case of a 2-player lottery, and b) a binary-tree tournament constructed from instances of the 2-player lottery. Both of our Bitcoin and Ethereum protocols require zero collateral and $O(\log N)$ rounds.

A surprising outcome of our work is thus that it points to a new separation between the expressive capability of the scripting systems in Bitcoin and Ethereum. Although fair lotteries have been a well-known demonstration of Bitcoin's versatility, our best Bitcoin-based protocol is significantly more expensive in terms of off-chain setup (i.e., it requires an off-chain setup phase with $O(2^N)$ signed transactions). Furthermore, the zero-collateral lottery is straightforward to implement in Ethereum's smart contract language, which easily expresses state machines that users update by sending transactions, whereas the Bitcoin version requires an unwieldy and indirect way of representing the state machine. Our evaluation (cf. Table 2) shows that the overhead costs are sensitive to the choice of language design, hence lotteries may continue to serve as a useful expressiveness benchmark. In Section 7, we propose a minimal modification to Bitcoin that would close the performance gap.

Although fair lotteries have been a well-known demonstration of Bitcoin's versatility, our best Bitcoin-based protocol is significantly more expensive in terms of off-chain setup (i.e., it requires an off-chain setup phase with $O(2^N)$ signed transactions). Furthermore, the zero-collateral lottery is straightforward to implement in Ethereum's smart contract language, which easily expresses state machines that users update by sending transactions, whereas the Bitcoin version requires an unwieldy and indirect way of representing the state machine. Our comprehensive evaluation (Table 2) shows that the collateral requirement is sensitive to the choice of language design, and thus may continue to serve as a useful expressiveness benchmark. In the Appendix, we propose a minimal modification to Bitcoin that would close the performance gap.

## 1.1 Background

**The Blockchain Model of computation** Our focus in this paper will be mostly concerned with ways in which Bitcoin [33] and Ethereum [41], two leading cryptocurrencies, differ. However, both systems are based on the Nakamoto "blockchain" protocol, and share mostly the same essential functionality as we review now. The underlying blockchain protocol can be thought of as a globally-ordered broadcast primitive, which can be configured for general purpose applications just by interpreting the contents of that log. In Bitcoin, the log is interpreted as a ledger of transactions that move quantities of money from one pseudonymous account to another. In Ethereum, the log is interpreted as a sequence of input commands to processes called "smart contracts" residing in a virtual operating system. Both systems provide some means of user programmability, allowing the flow of virtual money to be be constrained by fragments of code in a scripting language.

Ethereum provides a programming language resembling an actor-based process calculus — each "smart contract" process is defined as a behavior that reacts to each input and potentially modifies a local state or sends messages to other processes. Although this language is fairly general, the environment is well-known to be error prone [31, 15]. In principle, the language is Turing complete, however execution times are limited by transaction fees, called "gas", which are charged by the instruction. In contrast, the Bitcoin scripting language is mainly focused on signature verification, and can express a policy for accepting or rejecting a transaction, but cannot alter the effect of that transaction if accepted. A significant number of research papers have explored the possibility of building applications on top of cryptocurrencies, either by modifying the protocol [7, 29], or by using Bitcoin's [13, 10, 21, 38] or Ethereum's [15] existing scripting languages. A goal of our work is to better understand how the design of each language affects its usefulness by focusing on a particular task, the N-player lottery.

As is typical for modelling cryptocurrencies [18, 34, 35, 30, 24, 23], we assume that the underlying synchronicity of the communication network is such that transactions submitted to the blockchain system are accepted within a bounded time and that timestamp information available to the smart contract is approximately synchronized.

**Transaction Malleability and SegWit.** Our Bitcoin transaction construction relies on an upgrade to Bitcoin, called "Segregated Witness" (SegWit) [3, 17], which addresses a well-known problem of "transaction malleability" [5, 14]. At the time of this writing, SegWit has been developed, tested, and is pending deployment as a "soft-fork" upgrade.

In the current most commonly used version of the Bitcoin protocol, signature-scripts are considered to be part of the transaction-ID (TXID). This fact has the following unfortunate implication: given one signature-script for a transaction Tx, it is easy to construct different, but equally valid, signature-scripts. One may, for example, append an OP_NULL op code (which is simply skipped over, having no effect on transaction validation), resulting in a different signature script and therefore a different TXID. The ability to see a transaction Tx and derive a distinct transaction Tx' with a distinct TXID is called "transaction malleability." The transaction malleability problem is well-known, and prevents a variety of useful applications of Bitcoin script programs (smart contracts) based on transaction scaffoldings. Essentially, the transactions in a scaffolding construction are linked to each other by reference to their TXIDs; transaction malleability means that an attacker could direct their coins "outside" the scaffolding by crafting transactions with malleated TXIDs.

The SegWit upgrade eliminates transaction malleability by removing the signature script from the body of the transaction, and thus the new TXID format (a.k.a. NTXID) does not depend directly on the signature script. See [3, Section 3] and [25, Section 5.1] for more information.

Our transactions also rely on a recently-added script instruction (as of version 0.11.2 [40]), called CHECKLOCKTIMEVERIFY, which is used to specify a minimum time before which an output cannot be spent.

## 1.2 Related Work

Andrychowicz et al. [4] developed fair lotteries on top of the Bitcoin scripting language. However, to achieve the notion of fair lottery, they require the participants to place an additional security deposit of $O(N^2)$ coins, along with their actual bet of 1 coin each. Remarkably, [4] is a constant-round protocol, though it requires $O(N^2)$ on-chain transactions in total. Bentov and Kumaresan presented an alternative lottery [9], that requires $O(N)$ on-chain transactions but $O(N)$ rounds, and still $O(N^2)$ collateral. In these, the entire collateral must be held in deposit for maximum duration of the protocol. Hence, if we consider the measure rounds × collateral as the *time-value* cost of participation in the lottery (per player), then [4] has $O(N^2)$ time-value cost, [9] has $O(N^3)$ time-value cost, and our lotteries have $O(\log N)$ time-value cost.

In Ethereum, Delmolino et al. implemented 2-player commitment-based lotteries [15]. Our 2-player Ethereum lottery functions similar to theirs, though the extension to $N$ players with zero collateral is first described in our work.

Concurrently and independently of our work, Bartoletti and Zunino developed a Bitcoin lottery protocol [6] which shares the same underlying insight as ours (i.e., theirs is also based on a binary tournament tree comprising 2-player lottery instances). However, our Bitcoin-based protocol has several differences compared to theirs: first, their protocol relies on an additional new custom opcode, whereas our relies only on the pending SegWit upgrade; second, their lottery requires a constant $O(1)$ collateral, whereas as ours requires zero; third, their payouts distribution is non-ideal (i.e., an honest player is not guarnteed to win the pot with at least $1/N$ probability). However, the tradeoff of working within the constraints of Bitcoin(+SegWit) is that we require an exponential amount of off-chain work, whereas their protocol requires only $O(N^2)$. In Section 7, we propose an alternative opcode, MULTIINPUT, that also results in $O(N^2)$ off-chain work. In Table 2 we compare between these aforementioned lotteries and our lottery protocols.

As an alternative to commitment-based lotteries, one may rely on a Proof of Work (PoW) based beacon. That is, the lottery can use the hash of the blocks themselves as a public randomness source. The rationale is that it is difficult to influence this hash value, other than by discarding mining work (and thus foregoing the mining reward), see [2, 11, 36, 26, 8]. A secure implementation of this approach should prevent the winner from spending her prize before a significant number of PoW confirmations occur, as otherwise a chain reorganization will invalidate every transaction that depends on the prize (similarly to the coinbase transaction in Bitcoin). While the beacon approach is indeed popular in Ethereum lotteries[15, 27], the GHOST [39] variant of Ethereum makes it significantly easier to influence the identifier of the puzzle. If the miner of the last block does not like the outcome of the lottery, she can withhold her block temporarily, so that it later can be included as an "uncle block" which earns 80% of the original block reward. This way, a miner can influence the beacon without forfeiting the entire block reward. A more secure beacon would derive the output from multiple blocks [8], possibly including such uncle blocks.

## 2 Overview of our N-Player Commitment Lottery Protocol

Though we present multiple variations of a lottery protocol in different cryptocurrency platforms, we evaluate their security properties in a common framework, which we describe informally below. We then explain the common high-level structure of our protocols.

**Security properties of a Lottery.** We consider a standard network model concerning a set of $N$ players, some of which may be "corrupted" and controlled by an adversary. We assume the network includes a "blockchain," which keeps track of the amount of digital currency (denoted with $\text{\textbedbeth}$) owned by each player.

A protocol that implements the N-party coin-flip lottery should satisfy the following security properties:

- (Ideal case.) If all parties are honest, then each party earns $\text{\textbedbeth}(-1)$ with probability $1 - \frac{1}{N}$, and earns $\text{\textbedbeth}(N-1)$ with probability $\frac{1}{N}$.

- (Terminates in finite time.) If any party is honest, then the protocol completes by some time parameter $T_{\text{final}}$, after which all payouts have been completed.

- (Commit/Abort decision is made even sooner.) Our protocols guarantee that by a shorter time, $T_{\text{commit}}$, the protocol has either "aborted" or "committed". If the protocol has "aborted", then each honest party immediately receives a net payout of 0 (i.e., any deposited money is returned by time $T_{\text{commit}}$).

- (Stochastically Dominant Payoff.) If the protocol has committed (rather than aborted), then each party receives a payoff distribution $P'$ (by the time $T_{\text{final}}$), a payoff distribution $P'$ that stochastically dominates $P_{\text{ideal}}$. Stochastic dominance is a natural relation that guarantees. For value $\$X$, you are at least as likely in $P'$ as you are in $P_{\text{ideal}}$ to receive a payoff of $\$X$ or more.

- (Minimal Collateral.) Although the above properties constrain the *net* payout offered to each party, the protocol may require parties to deposit additional money up front (as a collateral deposit), which honest parties are guaranteed to receive back.

The properties above are only required to hold except for negligible probability. Particularly, the protocol relies on cryptography with a security parameter $1^\lambda$, and so "negligible probability" here means a negligible function of $\lambda$, e.g., $\texttt{negl}(\lambda) = 2^{-\lambda}$.

**The 2-party commitment-based lottery.** In a two player commitment-based lottery, the two players, Alice and Bob, commit to a random value $x$ and $y$ respectively. After both commitments are registered, the players open their commitments. The revealed values are combined to determine a pseudorandom bit (e.g., via $x$ `XOR` $y$ mod 2), such that if either player chooses a random value then the output is truly random. If, instead, either player fails to commit, or fails to reveal their committed value, then that player is punished (i.e., that player automatically forfeits her initial bet and may not win the lottery anymore).

Typically, the commitment is implemented using a hash function. A couple of caveats bear mentioning. First, the hash function should not be directly applied to a string in a small value space (e.g., a 2-party lottery implemented according to "Rock-Paper-Scissors" rules). If the search space is small, the "hiding" property is not achieved, since an attacker could brute force to learn the committed value. Second, to prevent "replay" attacks, the committed value can require some prefix. This would prevent Alice from replaying Bob's commitment, in which case Alice always wins since the `XOR`'d decommitments produce 0.

**From 2-Player to N-Player using a Lottery Tournament.** To form a lottery for an arbitrary number of players, the 2-player lottery can be composed in a binary-tree tournament structure. Assume (for now) that the number of players, $N$, is a power of two. In the first level of the tournament (level $\ell = 0$), each pair of players $2i$ and $2i + 1$ is pitted against each other in $N/2$ initial instances of the 2-player lottery, called matches ($\mathsf{match}_{\ell,i}$). Next, the winners of each level-0 lottery are pitted against each other, again in pairs. The process continues, until the final level $\ell = \log_2 N$, which determines the overall winner. See Fig. 4 for an illustration.

**Security of the Lottery Tournament.** An honest participant in the tournament protocol will win the ₿$N$ pot with at least $\frac{1}{N} - \mathtt{negl}(\lambda)$ probability (where $\lambda$ is the security parameter), satisfying the stochastic dominance property. This is because the honest player participates in up to $\log N$ competitions, and wins each of these competitions with at least $\frac{1}{2} - \mathtt{negl}(\lambda)$ probability. This is because an honest player will reveal a decommitment to a fresh secret, and will proceed to the next round of the tournament either because the decommitment of the competing player loses, or because the competing player timed out (cf. Fig. 5 for the Bitcoin lottery and `getWinner` in Fig. 1 for the Ethereum lottery). The success probability of the honest party is adjusted by a negligible $\mathtt{negl}(\lambda)$ amount, due to the possibility that a malicious player could break the hiding or binding properties of the commitment scheme.

As with the blockchain-based protocol, the tournament protocol needs to be secure against history-reversal attacks. For example, if two malicious players $P_1, P_2$ compete and $P_2$ wins, and then in the next 2-party lottery between $P_2$ and an honest player $P_3$ it turns out that the decommitment of $P_3$ wins against $P_2$ but would have lost against $P_1$, then $P_1, P_2$ must not be allowed to reverse to blockchain history so that $P_1$ wins their competition. To achieve this, the timeout values (using `CHECKLOCKTIMEVERIFY` in the Bitcoin variant) of the tournament transactions are set according to a confidence parameter $\tau$. For example, $\tau = 6$ implies a security level of 6 PoW confirmation.

## 3 Lottery in the Ethereum Model

In this section we describe an implementation of a commitment-based lottery in Ethereum, following the high-level plan described previously.

**Implementing the 2-party logic in an Ethereum smart contract.** The implementation of our two-player lottery component, shown in Fig. 1, is written in Ethereum's high-level "Serpent" programming language. A Serpent program consists of data declarations and top-level functions. If no explicit type annotation is given, data fields are assumed to be of the default 256-bit signed integer type. Users interact with a smart contract by publishing a transaction containing a procedure call, including the address of the contract, the name of the function to invoke, and arguments to pass. When a user creates a transaction, they must pay a "gas" fee, which roughly corresponds to the computational cost (e.g., storage plus number of opcodes) of executing the smart contract code.

Like many Ethereum smart contracts, our lottery contract is structured as a time-based state machine, where hardcoded deadlines indicating the transition time between states (expressed as a number of blocks, read from the `block.number` register). Within the time interval from T0 to T1 (which is a parameter $\tau$), the two participants in the 2-party lottery must submit their commitments, and within T1 to T2 they must open their commitments. After T2, the winner is determined, based on the opened values; if one player fails to complete their duty, then the other player wins by default.

Our implementation includes initialization routines for configuring each of the declared data fields. We omit the code listing for these routines for brevity, but explain the main idea here. The parties that play in the two-player contract can be specified in one of three ways. First, if the addresses for players Alice and Bob are initialized to a non-zero value, then those are the players that will participate in the game (in this case, lines 16-21 in Fig. 1 will have no effect). Alternatively, the players can be determined at runtime, by querying a method of another contract (either `left`, `right`, or `deposit`), as explained shortly.

The players interact with the 2-party contract through the `commit` and `open` methods. The `commit` method stores the player's commitment in the persistent storage array, `self.commitments`, which is indexed by each player address. The player's address is implicitly passed as an argument, in the `msg.sender` register, which always refers the party that invoked this method. In order to prevent replay attacks, the player's public key is required as a prefix of the opened value (See line 28 in Fig. 1).

**Assembling a Tournament Tree.** For assembling the two-player Ethereum lottery into a multi-player tournament, we need to ensure that the two players in each match are determined at runtime, based on the winners of matches in the previous level (indicated by the `left` and `right` data fields). The approach we take in Fig. 2 is simply to construct all the tournaments at once, and to initialize the contracts in level $\ell + 1$ with the addresses of the contracts at level $\ell$. An overall illustration of the contracts tree is given in Fig. 4.

**Implementing the deposit and payout mechanism.** In order to accept monetary deposits, and to determine the players in the first level ($\ell = 0$) lotteries, we construct a separate "master contract," listed in Fig. 3. Before making any deposit, each player checks that the system of 2-player lotteries is correctly formed, i.e., that the lottery each level is configured with a correct deadline times and references the correct lottery contract in the previous level. Finally, the players each make a transaction that deposits money into the lottery contract. If more than $N$ parties submit transactions concurrently, all but the first $N$ are canceled with no effect.

**Security and Performance Analysis.** First consider the two-party contract (Fig. 1), assuming that the party addresses `alice` and `bob` have been hardcoded. The `getWinner` method can only be called after time $T0$, and afterwards it can only take one value (since the fields it depends on, `self.commits` and `self.openings`, cannot be modified after $T2$). If either of the parties `alice` and `bob` is honest, then they must call the `commit` method before time $T1$ and `open` before $T2$, with a randomly chosen opening value. Because of the abort-handling conditions (lines 34-40), an honest party is either chosen as the winner by default, or else determined by the opening values of both players. In the case that the openings are both revealed, an honest party will win the lottery with at least $\frac{1}{2} - \mathtt{negl}(\lambda)$ probability, where $\lambda$ is the security parameter (due to the binding and hiding properties of the commitment scheme).

To assemble a tournament, some party must initialize a system of smart contracts. Python code for creating these contracts in shown in Fig. 2. In the tournament construction, the players' addresses are determined at runtime rather than being hardcoded. If both parties who are *supposed* to play in a 2-player lottery fail to invoke the `commit` method, then it is possible that these will both be set to a null value, in which case `getWinner` will return a null value for the winner. Regardless, in the tournament construction, if an honest party wins a lottery in level $\ell$ by time $T2$, then they will be considered a player by time $T0$ in a lottery at the level $\ell + 1$.

In the first level, the players in the lottery are determined by reference to the "Master contract" (see Fig. 3). This guarantees that at time $T_{\mathsf{Commit}}$, either `getPlayers` contains $N$ addresses including any honest party who deposited money, or else every party that deposited can immediately withdraw their coins. Otherwise, the payment can be withdrawn after time $T_{\mathsf{Final}} = O(\log N)$ according to the winner of the final tournament.

The asymptotic performance of the Ethereum lottery is summarized in Table 2. The Ethereum lottery does not require any collateral beyond the bet. Each transaction (i.e., invocation of `commit`, `open`, `deposit`, or `withdraw`) is of a constant size, requiring only a single signature from one party. Asymptotically, our Ethereum-based lottery achieves the best performance by each metric.

**Gas Costs in Ethereum.** An implementation of the Ethereum tournament is available at https://github.com/amiller/zero-collateral-lottery. The implementation consists of Serpent code for the 2-party lottery, as well a `pyethereum` simulation that constructs and implements the private code for each player.

```
1   data alice:addr
2   data bob:addr
3   data openings[]:uint256
4   data commits[]:uint256
5   data T0
6   data T1
7   data T2
8   data isFirstLevel
9   data index
10  data deposit:addr
11  data left:addr
12  data right:addr
13
14  def commit(c:uint256):
15      assert(self.T0 < block.number < self.T1)
16      if self.isFirstLevel:
17          if self.alice == 0: self.alice = self.deposit.getPlayer(2*self.index)
18          if self.bob   == 0: self.bob   = self.deposit.getPlayer(2*self.index+1)
19      else:
20          if self.alice == 0: self.alice = self.left .getWinner()
21          if self.bob   == 0: self.bob   = self.right.getWinner()
22      assert(self.alice == msg.sender or self.bob == msg.sender)
23      self.commits[msg.sender] = c
24      log(type=Commit, msg.sender, c)
25
26  def open(s:uint256):
27      assert(self.T1 < block.number < self.T2)
28      assert(sha3([msg.sender, s], items=2) == self.commits[msg.sender])
29      self.openings[msg.sender] = s
30
31  def getWinner():
32      assert(self.T2 < block.number)
33
34      # Timed out before T1
35      if self.commits[self.alice] == 0: return(self.bob:uint256)
36      if self.commits[self.bob  ] == 0: return(self.alice:uint256)
37
38      # Timed out before T2
39      if self.openings[self.alice] == 0: return(self.bob:uint256)
40      if self.openings[self.bob  ] == 0: return(self.alice:uint256)
41
42      # Ordinary case
43      x = (self.openings[self.alice] xor self.openings[self.bob])
44      if mod(x, 2) == 0: return(self.alice)
45      else: return(self.bob)
```

Figure 1: Implementation of the 2-Player lottery in Ethereum. From the time period T0 to T1, players may place commitments. In the next phase, from time T1 to T2, players may open their commitments. After T2, the `getWinner` method determines the winner, accounting for any timeouts. The players Alice and Bob can either be represented by hardcoded addresses, or else they can be dynamically determined (after time T0) by querying the `getWinner` method of the `left` and `right` subcontracts.

```python
def buildTree(players, $T_{\text{Commit}}$):
    contracts = {}
    contracts[0] = {}
    for i in range(N/2):
        contracts[0][i] = 2PContract(firstLevel=True, index=i,
                                    T0=TCommit)

    for L in range(1,log2(N)):
        contracts[L] = {}
        for i in range(2**(levels - 1 - L)):
            contracts[L][i] = 2PContract(left=contracts[L-1][2*i].address,
                                         right=contracts[L-1][2*i+1].address,
                                         T0 = TCommit + 2*$\tau$*L)
```

Figure 2: Python code for assembling an Ethereum N-Player Lottery tournament.

```
data players[]:addr
data n_players
data N
data complete
data deposits[]
data finaltournament:addr
data $T-$Commit)
data $T_{\text{Final}}$)

def getPlayers(i):
    return self.players[i]

def deposit():
    assert(msg.value == 1)
    assert(self.n_players <= self.N)
    assert(self.deposits[msg.sender] == 0)
    assert(block.number <= $T_{\text{Commit}}$)
    self.players[self.n_players] = msg.sender
    self.n_players += 1
    self.deposits[msg.sender] = 1
    if contract.balance >= N:
        self.complete = 1

def withdraw():
    if !self.complete:
        assert(block.number >= $T_{\text{Commit}}$)
        assert(self.deposits[msg.sender] == 1)
        self.deposits[msg.sender] = 0
        send(msg.sender, 1)
    else:
        assert(block.number >= $T_{\text{Final}}$)
        assert(finaltournament.getWinner() == msg.sender)
        send(msg.sender, contract.balance)
```

Figure 3: Serpent code for the "master contract" that accepts deposits and execute payments for the players in an N-player lottery. The master contract also refunds partial deposits in case of an abort before the lottery begins.

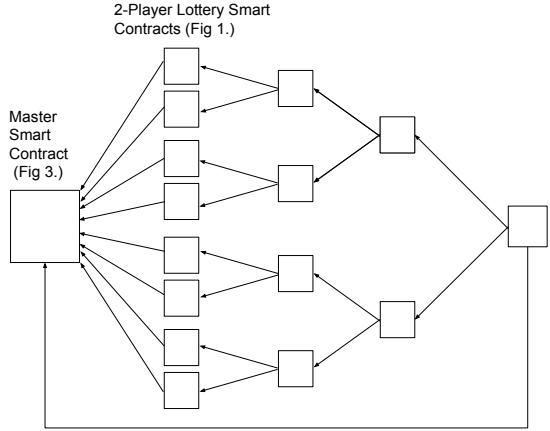

Figure 4: Illustration of the system of Ethereum smart contracts composing a tournament tree. The python code in Fig. 2 can be executed by any party in order to initialize the structure of contracts depicted here.

We ran experiments to determine the gas costs of our protocol (i.e., the transaction fees. Gas costs in Ethereum are determined by several factors. First, there is a fixed mapping for each instruction to "gas units," as defined in the Ethereum "yellow paper" reference document [41]. Second, miners enforce a minimum Ether price per gas unit, which today is approximately 2.5E-8 Ether per gas unit. Finally, the market price of Ethereum today is approximately $10 per Ether. Putting these together allows us to express the actual cost of our smart contract in dollars. In Table 1, we show the worst-case costs for each step of the 2-player lottery implementation (Fig. 1). Initialization is most expensive, though it needs to be performed only once. Each operation is independent of the number of parties $N$, however several of the operations must be performed $O(N)$ times. In total, the cost of a lottery is $-3.6 + 12.7N$ cents, e.g., an 8 player lottery would cost about a dollar in transaction fees today.

Table 1: Gas Costs for the N-Player Lottery in Ethereum (Figs. 1,2,3)

| Transaction | Gas | (USD) | # required |
|---:|---:|---:|---:|
| Initialize Master | 300012 | 7.5 cents | 1 |
| Initialize 2-Player | 181539 | 4.5 cents | $N-1$ |
| Deposit | 67270 | 1.6 cents | $N$ |
| Commit | 89812 | 2.2 cents | $2(N-1)$ |
| Open | 45706 | 1.1 cents | $2(N-1)$ |

## 4 Lottery in the Bitcoin Model

We present here a variant of the tournament protocol that can be implemented in Bitcoin (once the SegWit upgrade becomes operational). The only aspect of SegWit that our protocol relies upon is the use of NTXID instead of TXID to reference transaction inputs. In general, this feature enables smart contracts, in the following sense: the protocol can have multiple prepared transactions such that some of them may become valid in the future, depending on the way that a chain of prior transactions branched [28, 5].

This technique can be regarded as simulating covenants (cf. [32]). In the particular case of the lottery, every prepared transaction will require the signatures of *all* the players (and possibly some additional witnesses), and all players sign these prepared transactions before the on-chain protocol starts. Thus, if there is a single honest player then each transaction can be spent only via the prepared transactions that were already signed in advance.

**Transaction Scaffolding** . Our protocol consists of a transaction construction, as illustrated in Fig. 6. The protocol begins with a setup-phase, where the players assemble a "scaffold" of partially-filled out (but still incomplete) Bitcoin transactions that depend on one-another. The main phase of the protocol begins after all of the players have signed each incomplete transaction in the scaffold. Once signed, the transactions are still not complete. During the protocol, players take turns revealing secret information (previously committed-to in the scaffolding). The secret information, combined with each set of signatures, completes the "witness" portion of the transaction, thus enabling it to be incorporated into the blockchain.

We now describe the transaction structures that we need and conventions for composing them.

**The N-N Master Key.** Before the protocol begins its main phase, we require that each transaction in the scaffolding must be signed by *every one* of the $N$ players. We can implement this constraint in a Bitcoin script using the fragment [CHECKMULTISIGVERIFY{N}{N}{pubkeys}], which can be satisfied only by providing all the required signatures. However, this results in an on-chain cost of $O(\lambda N)$ per transaction. In Section 6 we describe how to reduce this cost using threshold or aggregatable signatures; the performance of these extensions are reflected in the "Lottery w/ Threshold" row in Table 2.

**Connecting Inputs to Outputs.** Each transaction input in Bitcoin consists of a reference to an output of a previous transaction, for example: Tx2.inputLeft = Tx1.output. These references are concretely represented as the hash of the previous transaction and the index of the output (in case the transaction has multiple outputs to disambiguate).

We construct the overall transaction scaffold by assembling groups of incomplete transactions called "kernels." Our two-player lottery shown in Fig. 5 is based on a kernel of two transactions, Tx1 and Tx2. The Tx1 transaction must be committed before the beginning of the lottery, T0, and references inputs that correspond to the two players. This transaction contains a commitment to a value from the left player (Alice).

There are three possible outcomes to the 2-player lottery, represented by the red boxes. In the first case, if Alice fails to open her commitment, then Bob can spend the Tx1 output, winning by default. Otherwise, Alice must spend Tx1 until time T1 by revealing her commitment and completing Tx2. Once Tx2 is committed, we coalesce the cases where Bob aborts versus when Bob loses fairly. Since Bob moves last, if he realizes that $a \oplus b = 0 \mod 2$ and hence he cannot win, he simply does not post any transaction, and the coin can be spent by any (pre-signed) transaction after time T2. Otherwise, if Bob can win, then the transaction can also be spent by a transaction that reveals his commitment.

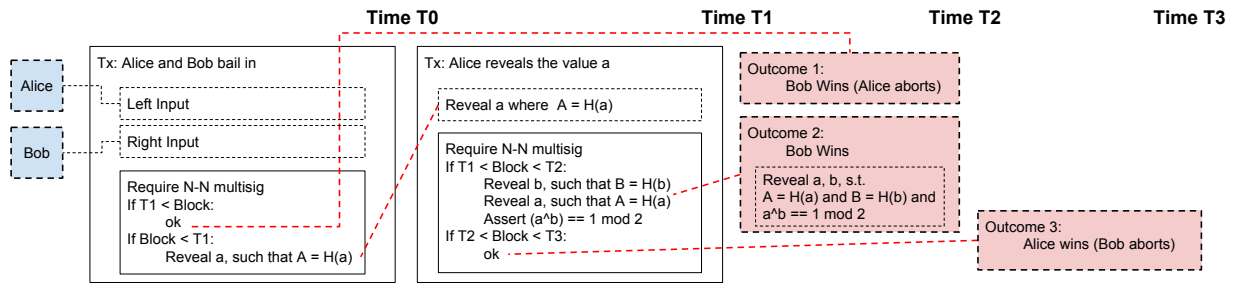

Figure 5: The 2-Player Lottery contract implemented using Bitcoin (with SegWit) transaction scripts. The contract requires two transaction inputs, one each for the left and right players, and can result in one of three possible outcomes. This is used as a "Kernel" to compose a tournament tree implementing an $N$-Player Lottery.

The Bitcoin rules prevent committing any transaction that "double spends" (i.e., refers to a transaction output that has already been spent once before). Thus, although our scaffolding contains multiple transactions that spend a common output, only one member of this set can be committed at runtime. This means that the three outcomes in each 2-party lottery are exclusive.

**Assembling the $N$-Player Tournament.** To construct a tournament out of the 2-player lottery described above, we need to overcome an inconvenient restriction in the Bitcoin transaction format, as we explain below.

Its natural to describe a tournament as a collection of $2N$ "matches," where the players in each match are determined by the winners of the matches at the previous level (i.e., $\mathsf{Match}_{\ell,i}$ is the $i^{th}$ game of the tournament at level $\ell$, where $\ell$ ranges from 0 to $\log_2 N$, and $i$ ranges from 0 to $2^\ell$). In the Ethereum lottery described in Section 3, we simply create one smart contract per match, and determine the players at runtime.

This approach does not work directly in Bitcoin. The problem is that each Bitcoin transaction needs to be constructed by including the transaction input references, before it can be signed. We end up needing one transaction for each *combination of outcomes leading up to a match*, rather than one per match.

We illustrate with an example. Consider a 4 party tournament. For the $\ell = 1$ level, the inputs clearly depend on the winners of the $i = 0$ and $i = 1$ matches at level $\ell = 0$. However, each of these tournaments results in one of 2 possible transaction outputs being committed onto the blockchain. Furthermore, the outputs may be spent in one of three possible ways, one where Alice wins and two where Bob wins. Since the transactions in a Kernel must explicitly identify the inputs that they can spend, there are a total of 3 x 3 possible combinations. The best we can do is to create a signed transaction for each of these 9 possibilities.

To extrapolate from this example to the general case, we essentially need to include a stack of kernels to represent each match. For a match in level $\ell + 1$, we need 9 kernels for each pairwise combination of kernels in the two preceding matches in level $\ell$. Ultimately, this requires a scaffolding with an exponential $O(2^N)$ number of transactions, even though only a linear $O(N)$ number need to be committed on the blockchain.

We define the number of kernels, $\#\mathsf{Kernels}_{\ell,i}$, necessary for each match in the tournament $\mathsf{Match}_{\ell,i}$, according to the following recurrence relation. $\#\mathsf{Kernels}_{0,i} = 1$, and $\#\mathsf{Kernels}_{\ell+1,i} = (3 \cdot \#\mathsf{Kernels}_{\ell,2i})(3 \cdot \#\mathsf{Kernels}_{\ell,2i+1})$. If we assume $N$ is a power of 2, then this recurrence relation is easily solved as $\#\mathsf{Kernels}_{\ell,i} = 9^\ell$. This recurrence leads directly to the asymptotic analysis of our protocol.

In Section 5, we provide an inductive definition for constructing each transaction in the scaffold, including the correct assignment of inputs to output references. We illustrate the construction in Fig. 6, using the graphical notation of a solid white sheet to represent a kernel, with internal transactions represented by small white squares, outputs represented by pink dashed squares, and input references represented by blue dashed squares.

**Implementing deposits and payouts.** We now describe the process by which players commit their deposits and receive the payout. First, the players must signal their willingness to play, and agree on an assignment of a player number to each identity. Each player $i$ generates a fresh keypair $(sk_i, pk_i)$ of secret and public keys, and her player ID is the public key $pk_i$. From the player IDs, they can construct an N-N threshold signature address, called the master key as described earlier. Next, the players create a transaction that commits 1 unit of BTC from each player. To ensure fairness, this can be done in two ways.

1. The players create a single unsigned atomic transaction whose inputs are their 1 BTC deposits and whose outputs will be the leaves of the tournament tree. After all the other transaction for the next levels of the tournament are signed, the players will finally also sign this atomic transaction. Recall that due to the use of NTXIDs in segwit, the unique transaction IDs for these $N$ outputs will remain the same after the atomic transaction is signed, and therefore the next transactions can depend on these NTXIDs. The drawback of this approach is that it imposes a limit on the maximal number of players $N$ that can participate in the lottery, since a fee for a very large transaction will be excessively high.[1]

2. Each player commits to the blockchain a deposit transaction that spends 1 BTC into a leaf of the tournament tree. However, two additional conditions need to hold. First, this deposit transaction uses `CHECKLOCKTIMEVERIFY` to allow the player to receive her money back after a long enough time in the future (after the *first round* of the tournament has completed), so that malicious players who refuse to sign all the next transactions of the tournament will not damage an honest player. Second, the output of the deposit transaction can be spent only if a preimage $\hat{x}$ such that $\hat{X} = \mathsf{hash}(\hat{x})$ is revealed, where $\hat{X}$ is hardcoded in the transaction. Prior to creating the deposit transactions, the players will execute a secure MPC where the input of each player is a random secret $x_i$, and the output is $\mathsf{hash}(x_1 \oplus x_2 \oplus \cdots \oplus x_N)$. After the deposits are committed on the blockchain and all the rest of tournament transactions are already signed, the players will reveal their secrets $x_i$ values. Since Bitcoin transactions are public, when any deposit is spent into a leaf of the tournament tree, the preimage $\hat{x} = x_1 \oplus x_2 \oplus \cdots \oplus x_N$ becomes known, which implies fairness.

---

[1] The practical limit for $N$ can still be good enough, see e.g. a transaction with 412 outputs: `https://bitcoinchain.com/block_explorer/block/283873/`.

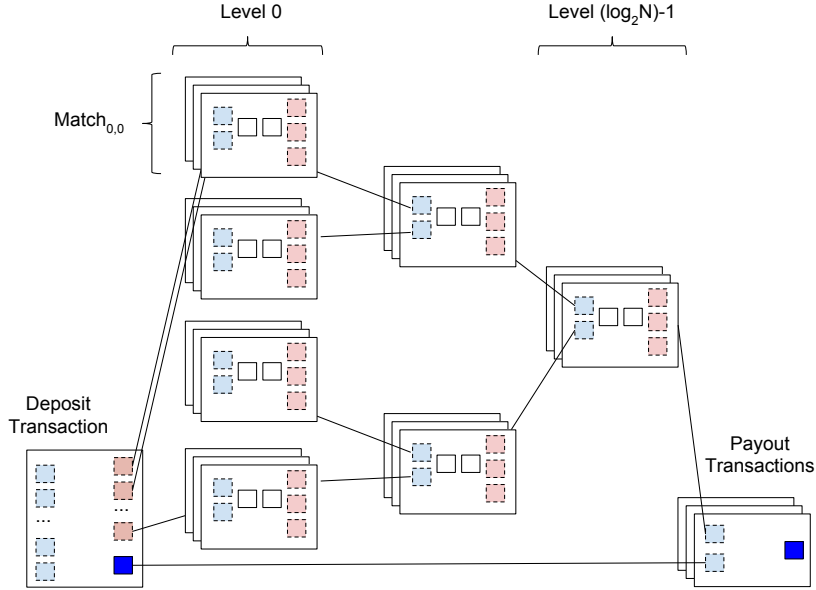

Figure 6: Illustration of the Tournament, implemented in Bitcoin. The tournament consists of of $2N-1$ matches. Each match consists of a set of kernels, one kernel for each combination of outcomes of the two matches leading into it in the tournament. The number of kernels in each match at level $\ell$ is $\#\mathsf{Kernels}(\ell) = 9^{\ell}$.

If the initial deposits phase is successful (i.e., before time T0), then the parties next assemble and sign the entire set of transactions according to the construction defined in Section 5. Finally, after receiving all of the N-N signatures for each transaction, the parties sign the atomic commit transaction described above (or reveal their MPC secrets, depending on which of the above options is taken).

## 5 Detailed description of the Bitcoin Tournament

To finish out the formal description of the protocol, we need to describe how the inputs and outputs of the transactions are hooked up, and how the transactions are configured. There are three "output" transactions in each kernel, TxA, TxB, and TxB', where each of these transactions have a single output. Since there are $\#\mathsf{Kernels}_{\ell,i}$ kernels associated with each match, we add another index $j$ to the Kernels for each configuration. To describe the connection relation between the inputs of each kernel to the output of other kernels, we essentially need to unpack the index $j$ to identify the corresponding input kernels, and the particular transaction in each kernel. Expressing this packing relationship is a bit tedious, and prone to off-by-one error. Recall that $j$ ranges from 0 to $\#\mathsf{Kernels}(\ell+1, i) = (3 \cdot \#\mathsf{Kernels}(\ell, 2i))(3 \cdot \#\mathsf{Kernels}(\ell, 2i+1))$. Thus we first unpack $j$ into components

$$\mathsf{IndexLeft}(\ell+1, i, j) = j/(3 \cdot \#\mathsf{Kernels}(\ell, 2i+1))$$
$$\mathsf{IndexRight}(\ell+1, i, j) = j \mod (3 \cdot \#\mathsf{Kernels}(\ell, 2i+1))$$

so that IndexLeft identifies the index of the left input kernel and transaction, and IndexRight determines the index of the right. These indexes are then unpacked as a kernel index, in the range 0 to $\#\mathsf{Kernels}(\ell, 2i+0, 1)$, and a "Tx" index, in the range 0 to 2.

$$\mathsf{IndexLeftKernel}(\ell, i, j) = \mathsf{IndexLeft}(\ell, i, j)/3$$
$$\mathsf{IndexRightKernel}(\ell, i, j) = \mathsf{IndexRight}(\ell, i, j)/3$$
$$\mathsf{IndexLeftTx}(\ell, i, j) = \mathsf{IndexLeft}(\ell, i, j) \mod 3$$
$$\mathsf{IndexRightTx}(\ell, i, j) = \mathsf{IndexRight}(\ell, i, j) \mod 3$$

This allows us to assemble the input and output connections for each of the kernels.

$$\mathsf{Kernel}_{\ell+1,i,j}.Tx0.\mathsf{inputLeft} = \mathsf{Kernel}_{\ell,2i,\mathsf{IndexLeftKernel}(\ell,i,j)}.Tx[\mathsf{IndexLeftTx}(\ell,i,j)].\mathsf{output}$$
$$\mathsf{Kernel}_{\ell,i,j}.Tx0.\mathsf{inputRight} = \mathsf{Kernel}_{\ell,2i+1,\mathsf{IndexRightKernel}(\ell,i,j)}.Tx[\mathsf{IndexRightTx}(\ell,i,j)].\mathsf{output}$$
$$\mathsf{Kernel}_{0,i,0}.Tx0.\mathsf{inputLeft} = \mathsf{DepositTx.output}[2i]$$
$$\mathsf{Kernel}_{0,i,0}.Tx0.\mathsf{inputRight} = \mathsf{DepositTx.outputs}[2i+1]$$

Each Kernel must also be instantiated with a random string known to each of the two players. Although the winning player in each match is determined at runtime through the execution of the protocol, the index of each Kernel already indicates a particular set of inputs associated with a particular path through the tournament, so we can define define the players for each kernel with the following.

$$\mathsf{Winner}(\ell, i, 3*j+0) = \mathsf{Kernel}_{\ell,i,j}.\mathsf{LeftPlayer}$$
$$\mathsf{Winner}(\ell, i, 3*j+1) = \mathsf{Kernel}_{\ell,i,j}.\mathsf{RightPlayer}$$
$$\mathsf{Winner}(\ell, i, 3*j+2) = \mathsf{Kernel}_{\ell,i,j}.\mathsf{LeftPlayer}$$
$$\mathsf{Kernel}_{0,i,0}.\mathsf{LeftPlayer} = \mathsf{Player}2i$$
$$\mathsf{Kernel}_{(0,i,0)}.\mathsf{RightPlayer} = \mathsf{Player}2i+1$$
$$\mathsf{Kernel}_{\ell+1,i,j}.\mathsf{LeftPlayer} = \mathsf{Winner}(\mathsf{IndexLeft}(\ell,i,j))$$
$$\mathsf{Kernel}_{\ell+1,i,j}.\mathsf{RightPlayer} = \mathsf{Winner}(\mathsf{IndexRight}(\ell,i,j))$$

To avoid "replay" attacks in which a malicious players picks the same commitment hash as an honest player so that the XOR'd decommitments will be 0, an honest player should refuse sign the initial deposit transaction (cf. Section 4) in the case that there exist two equal commitment hashes in the transactions of the tournament tree.

## 6 Reduced cost with Aggregatable or Threshold Signatures

To improve on our multisig-based Bitcoin lottery protocol, we can apply two approaches. The first is to use the Bitcoin-compatible threshold signatures of Goldfeder et al. [19], although distributing the initial keys to the players requires either a complex multiparty setup procedure with zero-knowledge proofs, or else a trusted dealer. An alternative approach is to make use of the planned support for Schnorr signatures in Bitcoin [16], as this implies an off-chain signature-aggregation protocol whose complexity is on par with the standard Bitcoin `CHECKMULTISIGVERIFY` opcode. In our analysis (Table 2), we include a row "Lottery w/ Threshold" that accounts for the improved on-chain cost when either of these approaches is taken.

Table 2: Performance Comparison of all Cryptocurrency Lottery Schemes. All values are worst-case. Off-chain bytes are per party.

| Scheme | Collateral | On-Chain | Off-Chain | Rounds $T_{\text{Commit}}$ | $T_{\text{Final}}$ | Distribution of Payouts | Requirements |
|---|---|---|---|---|---|---|---|
| ADMM14[4] | $O(N^2)$ | $O(N^2)$ | — | $O(1)$ | $O(1)$ | Ideal | Bitcoin |
| BK14[9] | $O(N^2)$ | $O(N^2)$ | — | $O(N)$ | $O(N)$ | Ideal | Bitcoin (+CLTV) |
| BZ16[6] | $O(1)$ | $O(N)$ | $O(N^2)$ | $O(\log N)$ | $O(\log N)$ | Non-ideal | Bitcoin (+Templates) |
| Bitcoin Lottery (Fig. 6) | 0 | $O(N^2)$ | $O(2^N)$ | $O(1)$ | $O(\log N)$ | Ideal | Bitcoin (+SegWit) |
| Bitcoin w/ Threshold (Section 4) | 0 | $O(N)$ | $O(2^N)$ | $O(1)$ | $O(\log N)$ | Ideal | Bitcoin (+Schnorr) |
| Bitcoin w/ (MULTIINPUT) | 0 | $O(N)$ | $O(N^2)$ | $O(1)$ | $O(\log N)$ | Ideal | Bitcoin (+MULTIINPUT) |
| Ethereum Lottery (Figs. 1,2,3) | 0 | $O(N)$ | — | $O(1)$ | $O(\log N)$ | Ideal | Ethereum |

## 7 MULTIINPUT Script Instruction

To avoid the exponential blowup in off-chain transactions, we propose to extend the Bitcoin scripting language by adding the MULTIINPUT opcode. Similarly to the standard CHECKSIGVERIFY, the MULTIINPUT opcode verifies the validity of a signature on the implicit transaction data. However, this implicit data now excludes the pair of input NTXID and index $(tx_0, i_0)$. Instead, MULTIINPUT takes a hardcoded set $S = \{(tx_j, i_j)\}_{j=1}^{c}$ of such pairs, and validates that the input $(tx_0, i_0)$ belongs to $S$. Thus, if a valid signature on the transaction is already available, any currently unspent output from $S$ can be attached to the transaction and thereby make it valid.

In the case of the tournament protocol, the unique outcomes of Fig. 5 across all the possible identities of the winners can be "compressed" via MULTIINPUT into a single outcome. For example, if player $P_2 \in \{P_1, P_2, P_3, P_4\}$ wins against either of $\{P_5, P_6, P_7, P_8\}$, then there are $c < 2 \cdot 4$ combinations of unique inputs that imply that $P_2$ won her competition in this level of the tournament, where the "2" factor is due to the different ways to win (cf. Section 4 and Fig. 5), and the "4" factor corresponds to the combinations $(P_2, P_5), (P_2, P_6), (P_2, P_7), (P_2, P_8)$. Hence, these $c$ unique combinations can be compressed via a MULTIINPUT transaction, and therefore there would be only a single NTXID that corresponds to $P_2$ winning in this level of the tournament.

Without MULTIINPUT, the total waiting time until the Bitcoin tournament terminates is $2\tau \cdot \log N$, where $\tau$ is the confidence level for irreversibility (e.g., $\tau = 6$ block confirmations) and the factor 2 is due to the intermediate extra transaction that is needed in Fig. 5. Therefore, with MULTIINPUT the total waiting time until termination becomes $4\tau \cdot \log N$, due to the extra "compression" transactions.

Since MULTIINPUT allows for unique NTXIDs as the identities of the winners at each 2-party lottery of the tournament, the overall amount of off-chain signed transactions is $O((\frac{n}{2})^2 + 2 \cdot (\frac{n}{4})^2 + 4 \cdot (\frac{n}{8})^2 + \ldots) = O(n^2)$.

## 8 Discussion and Conclusion

We have presented two protocol variations of fair commitment-based lotteries, implementable in Ethereum and in Bitcoin. Our protocols overcome the significant $O(N^2)$ "collateral cost" of prior-known lotteries, eliminating this requirement altogether. See Table 2 for a comparison between the prior lotteries and our lotteries. The improved lottery protocols that we presented can be deployed as applications in their own right (i.e., for a gambling game), or as a coin-flipping primitives for leader election (e.g., as part of a proof-of-stake protocol). We reflect on the implications of our work for the design of cryptocurrency platforms.

**Collateral Costs.** Based on our effort in optimizing the collateral requirements of the Bitcoin lottery, we conjecture that the $O(2^N)$ setup requirement *cannot* be overcome for zero-collateral lotteries in the Bitcoin(+SegWit) model, and that zero-collateral lotteries are impossible in Bitcoin at the time of writing (prior to SegWit).

In our analysis, we make a fine-grained evaluation of the collateral costs of the protocol, including both the magnitude (measured in dollars) and the time it is encumbered (e.g., we distinguish between early release of collateral in the abort phase and the final phase). However, many applications, not just lotteries, also rely on collateral deposits in order to enforce properties like fairness. We therefore recommend that collateral costs be explicitly considered when evaluating cryptocurrency applications more generally.

**Qualitative Advantages of the Ethereum Tournament over the Bitcoin Tournament.** In addition to the reduced complexity, the stateful nature of Ethereum transactions also provide resistance to denial of service (DoS) attacks. Indeed, our Ethereum lottery tournament contracts do not rely on hardcoded identities, and therefore any $N$ players that join the tournament (by making a valid deposit before the initial deposits timeout $T_{\mathsf{Commit}}$) will force the state of the lottery to become operational. The reason for this is that each valid deposit is published on the blockchain, and thus turns irreversible via PoW. The Bitcoin tournament of Section 4 can also be regarded as a free-for-all lottery, but it is susceptible to DoS. That is, any set of players can agree to run the off-chain protocol for signing the tournament transactions, starting from the root and progressing towards the leaves. However, a malicious player who would then abort instead of making the initial deposit will cause all the other players to restart (hence, the DoS is at least as severe as CoinJoin based mixing, cf. [37, 38, 22, 21]).

**Compositionality of Ethereum vs. Bitcoin.** Bitcoin and Ethereum both provide smart contract programming interfaces. However, the philosophy between the two are markedly different. Ethereum favors generality and expressiveness, while Bitcoin is intended to provide a small attack surface. While Bitcoin has gradually expands the expressiveness [5, 40, 17] of its language, Ethereum begins with a "Turing complete" language intended for general-purpose use. Many desirable functionalities, such as micropayments, joint accounts, kickstarter-style assurance contracts, can be implemented in either language [20, 12].

A virtue of the Ethereum language design is that contracts provide a nicely composable abstraction, which our Ethereum protocol makes use of. In fact, Ethereum's process-based model (each contract is effectively an independent process in the operating system) closely matches the pseudocode used in cryptography. [24, 30] Based on this experience, we make a general recommendation that when developing a new cryptocurrency application, one begins by using a generic Ethereum-model pseudocode, and then adapt it to a Bitcoin-specific implementation if desired.